\documentclass{eptcs}
\providecommand{\event}{CREST 2016} 
\usepackage{breakurl}             

\usepackage[english]{babel}
\usepackage[utf8]{inputenc}
\usepackage{amsmath}
\usepackage{xspace}
\usepackage{csquotes}
\usepackage[shortlabels,inline]{enumitem}
\usepackage{todonotes}
\usepackage{listings}
\usepackage{wrapfig}
\usepackage{placeins}
\usepackage{cleveref} 

\title{Towards a Unified Model of\\ Accountability Infrastructures}
\author{Severin Kacianka
\institute{Technische Universität München}
\institute{Garching b. München, Germany}
\email{kacianka@in.tum.de}
\and
Florian Kelbert
\institute{Technische Universität München}
\institute{Garching b. München, Germany}
\email{kelbert@in.tum.de}
\and 
Alexander Pretschner
\institute{Technische Universität München}
\institute{Garching b. München, Germany}
\email{pretschn@in.tum.de}
}
\def\titlerunning{Towards a Unified Model of Accountability Infrastructures}
\def\authorrunning{S. Kacianka, F. Kelbert \& A. Pretschner}

\newcommand{\Eg}{E.g.\@\xspace}
\newcommand{\eg}{e.g.\@\xspace}
\newcommand{\ie}{i.e.,\@\xspace}
\newcommand{\cf}{c.f.\@\xspace}

\newcommand{\etal}{et al.\@\xspace}

\begin{document}
\maketitle

\begin{abstract}
Accountability aims to provide explanations for why unwanted situations occurred, thus providing means to assign responsibility and liability. As such, accountability has slightly different meanings across the sciences. In computer science, our focus is on providing  explanations for technical systems, in particular if they interact with their physical environment using sensors and actuators and may do serious harm. Accountability is relevant when considering safety, security and privacy properties and we realize that all these incarnations are facets of the same core idea. Hence, in this paper we motivate and propose a model for accountability infrastructures that is expressive enough to capture all of these domains. At its core, this model leverages formal causality models from the literature in order to provide a solid reasoning framework. We show how this model can be instantiated for several real-world use cases.
\end{abstract}

\section{Introduction}
\label{sec:intro}
\label{sec:introduction}
\begin{wrapfigure}{R}{5cm}
  \centering
   \includegraphics[width=0.3\textwidth]{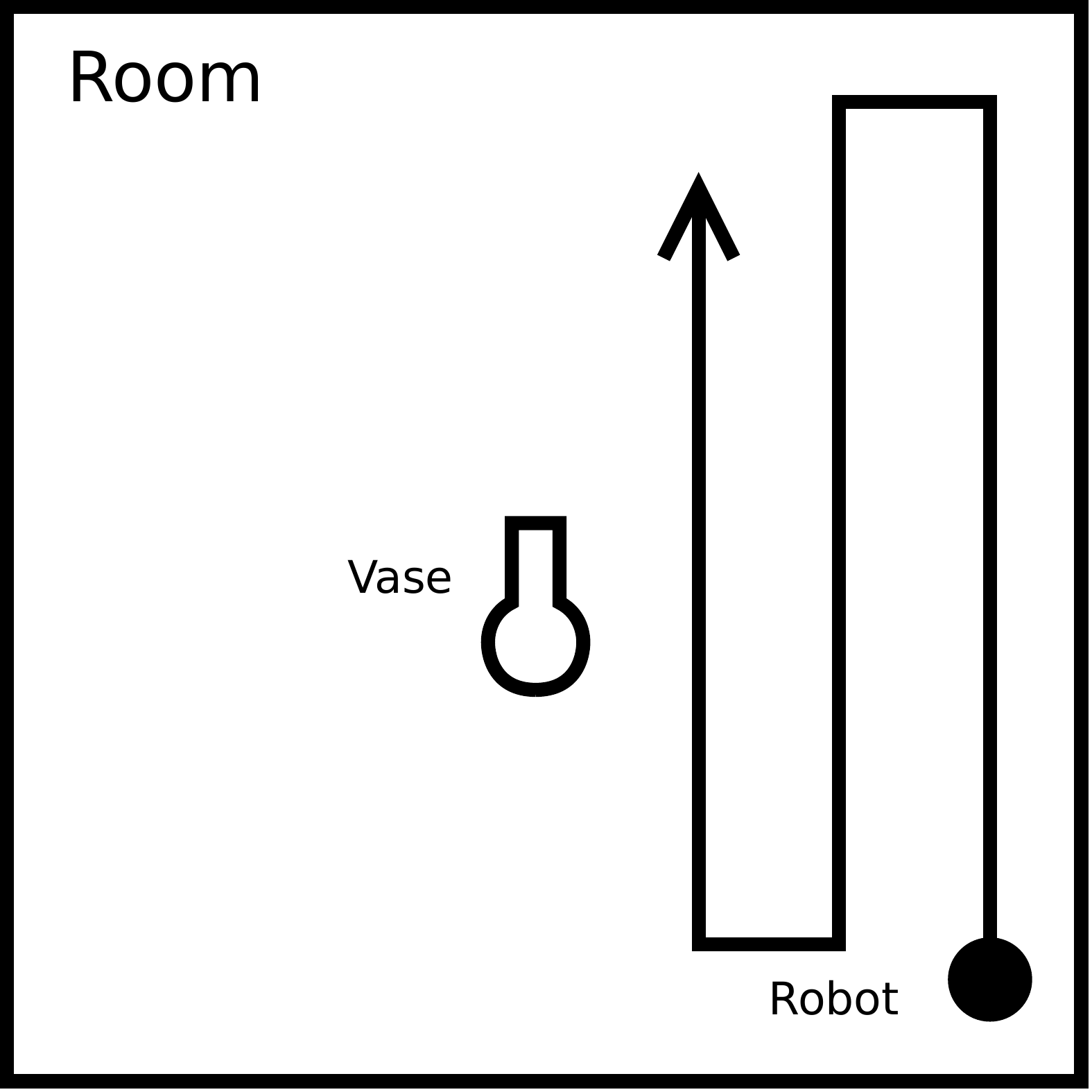}
  \caption{The vase room.} 
   \label{fig:vase_room}
   \vspace*{-1\baselineskip}
\end{wrapfigure}
We want to illustrate our understanding of accountability with a simple
scenario: Imagine a vase being placed in the middle of an otherwise empty room.
You place a cleaning robot (like the popular Roomba model) in the same room and
set it to clean the room. You leave the room for one hour, and upon return you
find the vase shattered; \Cref{fig:vase_room} sketches the scenario.

Given this, you would like to answer the following questions:
\begin{enumerate}
\item What caused the destruction of the vase?
\item Who is responsible for the destruction of the vase?
\item Who is liable for the destruction of the vase?
\end{enumerate}

At first glance, it might seem that the answer to these questions is obvious:
the robot. After careful consideration, however, the answer might not be so
clear. We have no clear \emph{evidence} of the robot's action. An alternative
explanation might be that a window was left open and that a cat snuck into the
room and tipped the vase. Another explanation could be an earthquake in the
area.

Thus, in order to answer the questions above, we need some \emph{evidence}
whether it was indeed the robot that caused the vase to be shattered.
\emph{Accountability mechanisms} are able to provide such evidence. While such
accountability mechanisms require technical components, they necessarily also
need non-technical components. To answer the question of liability, for
instance, lawyers need to be part of the overall system. Evidence cannot only be
provided by technical components, but also by humans (\eg, Susy saw that the cat
tipped the vase over) or knowledge of the world (\eg, there was an earthquake).

Generalizing from this example, we realize that accountability is relevant in
many domains. As the word suggests, it is relevant in financial accounting. It
is relevant when building road vehicles according to  ISO 26262, or when
building high security systems according to the common criteria. It is relevant
when running systems that are governed by standards like ITIL \cite{ITIL}, COBIT
\cite{COBIT}, or HIPAA \cite{HIPAA}.
In practice, accountability is implemented in airplanes with voice recorders, or
black boxes, according to JAR-OPS. While this list is not meant to be
comprehensive, it shows that accountability is comprehensive. It is concerned
with safety, security, and privacy, as well as the adherence to laws,
regulations, and standards. Understanding accountability may hence appear to be
a daunting task. We believe, however, that the conceptualization presented in
this work is sufficiently rich to capture all these domains. At the same time,
however, we are fully aware that in such complex environments  \enquote{We do
not know.} may sometimes be the best answer that an accountability mechanism can
provide.

In our considerations we focus on accountability for cyber-physical systems that
operate in the real world, and we pay attention to both technical and social
components. Indeed, from our point of view accountability is especially relevant
for systems that interact with humans.

\section{Related Work}
\label{sec:relwork}

\begin{wrapfigure}{R}{3.8cm}
  \centering
   \includegraphics[width=3.5cm]{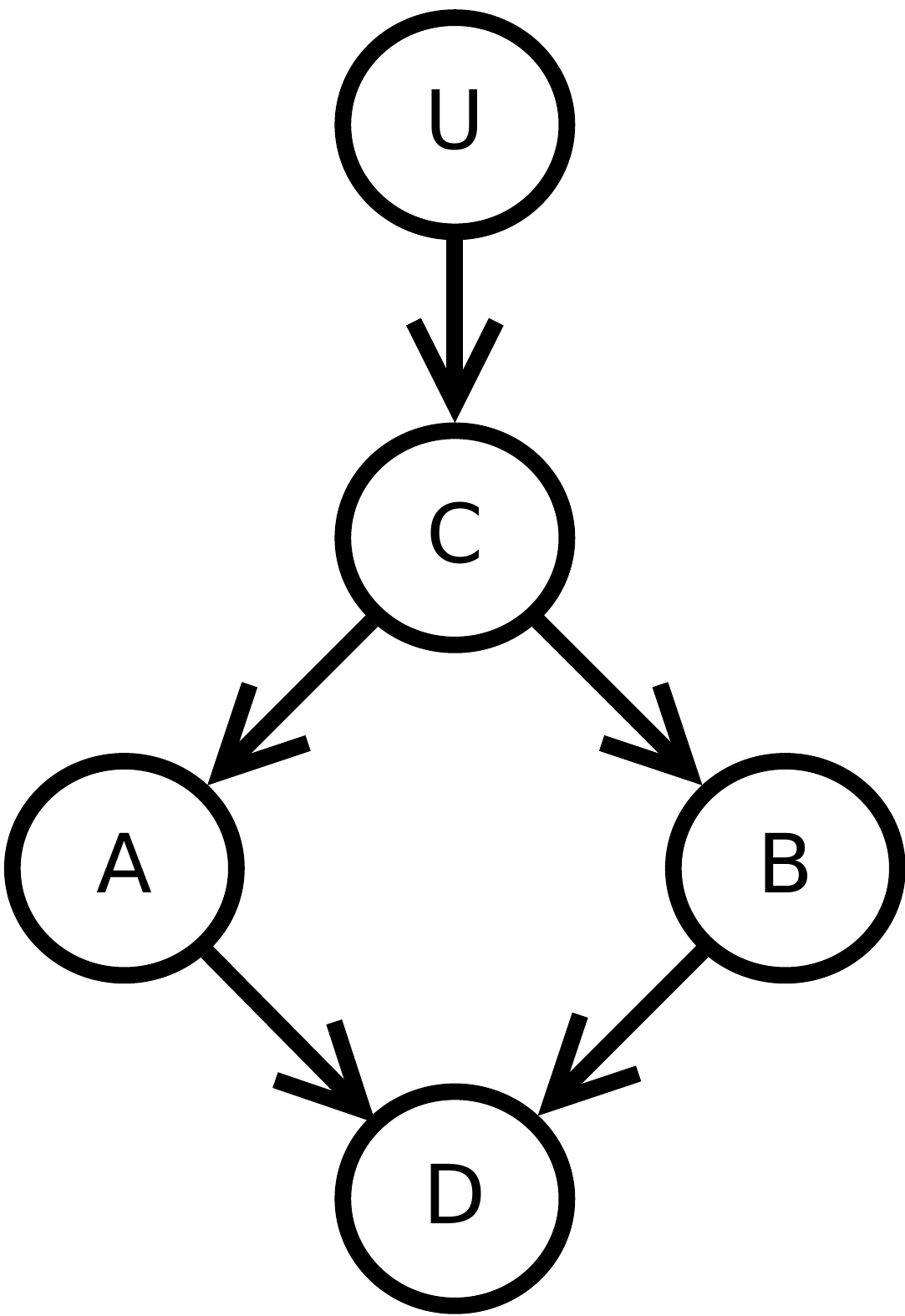}
  \caption{Firing squad example 
  \cite{pearl2009causality}.%
  }
   \label{fig:firing_squad}
\end{wrapfigure}

In computer science, the term \emph{accountability} was popularized by the seminal paper by Weitzner \etal \cite{weitzner2008information}. They introduced accountability as a way to ensure privacy and contrasted it with the traditional approach of achieving privacy through information hiding. In \Cref{sec:instantiation-privacy} we show how their work is an instance of our proposed model. Following Weitzner et al., research focused on accountability as a privacy mechanism and as a property of cryptographic protocols. K\"uster \etal \cite{kusters2010accountability} provide a definition of accountability that links accountability to verifiability. 
With the rise of cloud computing, privacy concerns were naturally applied to the cloud and accountability was seen as a way to ensure data security, \eg by Pearson et al. \cite{pearson2009accountability}. Suggesting accountability as a mechanism to ensure privacy was also the result of the PATS EU project and is detailed in a book edited by Guagnin \etal \cite{guagnin2012managing}. Xiao \etal \cite{xiao2012survey} provide a survey of accountability in computer networks and distributed systems. Finally, Papanikolaou and Pearson \cite{papanikolaou2013cross} summarize the understanding of the term accountability in different fields and contexts.

In contrast to aforementioned literature, we see accountability not merely as a mechanism to protect privacy or as an extension to non-repudiation. We believe that accountability aids us in
\begin{enumerate*}[(i)]
	\item understanding a system,  
    \item improving a system, and, most importantly,
    \item attributing actions of a system to responsible entities (people or organizations)
\end{enumerate*}. We further believe that accountability mechanisms can be generalized to be applicable to multiple domains. 

\medskip
\emph{Causality} was given a mathematical foundation by the works of  Pearl
\cite{pearl2009causality} and his collaboration with Halpern
\cite{halpern2005causes1,halpern2005causes2}, which resulted in the so-called
\enquote{Halpern-Pearl Definition of Causality}. Their definition originates in
their earlier work on Bayes networks (which can be also be used for inference,
learning and error diagnosis). Halpern published a modified version of their definition recently \cite{halpern2015}. They base their model of causality on \emph{counterfactuals}. The idea of counterfactuals is that an event A is the cause of another event B, if B would not have happened without A. 
While this definition is sufficient in many cases, there is a host of examples in the literature where it fails; see  Halpern \cite{halpern2015} for an overview. A comprehensive criticism of using counterfactuals for causal reasoning was published by Dawid \cite{dawid2000causality}. 

\Cref{fig:firing_squad} is the graphical representation of an example given by  Pearl  \cite{pearl2009causality}. In this example, the court (U) orders the execution of a prisoner. The order to shoot is given by a captain (C) and carried out by two rifleman (A and B). The final result is that the prisoner dies (D). Pearl and Halpern show how such a model can be constructed and used to reason about causes and possible alternative worlds. Their model of causality allows us to reason about sentences like \enquote{If the prisoner is dead, then the prisoner would be dead even if rifleman A had not shot.}

Recently, G\"ossler with Le M\'etayer and Astefanoaei respectively \cite{gossler2014general,gossler2014blaming} applied the idea of causal reasoning with counterfactuals to the fault analysis of real-time and embedded systems. They model system components as timed automata and use execution traces to blame individual components. 

Concepts similar to accountability can be found under different names, \eg fault localization or error tracing, and are often summarized under the term \emph{dependable computing}. An overview and a taxonomy of these terms was published by Avivzienis \etal \cite{avivzienis2004basic}.





\section{Preliminaries}
\label{sec:prelim}
\label{sec:preliminaries}

Before providing our model of accountability in \Cref{sec:model}, this \namecref{sec:prelim}  introduces key concepts that are central in the accountability context.

\subsection{Means to Provide Accountability}

Getting back to our example (\Cref{fig:vase_room}), we want to understand which means, both technical and non-technical, exist to find out the cause of the vase’s destruction. Later, the findings of any such means will be considered in the process of assigning responsibility.

\subsubsection{External Observation Systems}
\label{sec:prelim-external}
In general, an external observation system is a system that is passive regarding the events under investigation (\ie does not influence them). It merely records events in a log. We realize that in practice a definition of \enquote{external} necessitates a clear description of system boundaries. 

In our example, an external observation system providing accountability could be a person watching the room and telling us what has happened; or a video surveillance system that records everything that happens. However, both approaches might be insufficient: a person may lie (on their own accord or acting under duress), while the video system's recordings might be tampered with. None of these two means might be able to sense the light earthquake leading to the vase's destruction. Voice recorders in airplanes are another example for such an external observation system.

Another problem with external observation systems is that in many cases they are simply inexistent or hard to build. Hence, we can not just assume their existence. While software running on a computer, for example, can be monitored by the operating system, a cyber-physical system, such as a diving robot or a long-distance truck, will often act autonomously and without any possibility of direct observation.

\subsubsection{Internal System Logs}
\label{sec:prelim-logs}

To provide accountability, technical systems may keep some form of log. In such a log (or log book), all observable and (potentially) relevant events are recorded. Historically, logs for computer systems are collected in unstructured text files that seldom follow a common standard or convention. The log entries are usually driven by the needs of developers and administrators and as such they are not necessarily useful to understand the cause of a system’s behavior. 

In the vase room example (\cf \Cref{sec:intro}), the cleaning robot might have recorded a collision at a certain time, the reason being that one of the robot's bumpers (\cf \Cref{sec:platform}) was triggered and consequently a corresponding event was logged. However, with only this knowledge we could not tell if the robot collided with a wall (which is intended behavior) or with the vase. An expert might be able to analyze the logs and tell that collisions with walls are usually either around 30 seconds apart (the time it takes the robot to cross the room) or 5 seconds apart (when the robot navigates a turn). If, on the other hand, a collision is recorded after 15 seconds, then this observation could suggest that the robot bumped into something unexpected---likely the vase. This, however, may not be the only explanation. The robot could also have bumped into the cat, which in turn got scared and jumped into the vase. As a result, we realize that while recording the triggering of the robot's bumpers does provide some essential knowledge, this information might not be sufficient to reason comprehensively about the cause for the vase's destruction. Recording further information, such as the robot's position every five seconds, might have provided sufficient information.

To improve the utility of logs, we suggest that logging mechanisms used as part of an accountability mechanism should be both \emph{standardized} and \emph{extendable}. Standardization (\eg, using the same date format and the same taxonomy of events) allows for logs from different technical systems to be correlated and analyzed using different (standardized) tools. Extendability of logging mechanisms is another important requirement, because it will never be possible to enumerate all (potentially) relevant events at design time. \Eg, in our robot example the relevance of the robot's position might not have been considered beforehand. Hence, every cyber-physical system’s logging facility should be designed with extendability in mind. This is particularly important because the system's actual software might not be updated frequently and hence updating its logging mechanism out-of-band seems to be an adequate alternative approach. In the example at hand we could possibly further log the resistance that the bumper met. Colliding with a solid wall will likely cause a harder impact than colliding with a fragile vase (which will quickly give way) or the light touch required to startle a cat. Although we might not be able to find the cause (in an automated manner) the first time, an extendable logging facility can make sure that we can determine the cause in similar situations in the future.

Note that by definition it is challenging to imagine how unanticipated events can be handled at the design time of the system. Interestingly, combining internal system logs with external observation systems as described in \Cref{sec:prelim-external} might be able to address this problem, since the latter will in many cases also record certain types of unanticipated events.

\medskip
\noindent
The recordings of both, external observation systems and internal system logs, will be considered in the process of assigning responsibility. Thus, it seems natural to require these recordings to be sound (recorded facts have actually happened), complete (facts that have happened and that should be recorded, have actually been recorded), and tamper-proof (recordings should not be modifiable in hindsight, be it by fire, water, or malicious adversaries). In addition, we require the process of event recording to be efficient, meaning that the act of logging does not impair system usability.

\subsection{Assigning Responsibility}
The answer to the second question ---Who is responsible for breaking the vase?---  should name a person or entity that started the chain of events that ended with the vase breaking. Unfortunately, this answer cannot always be found and will often be ambiguous. If the vase was shattered by an earthquake, no one is directly responsible. If the cat entered the room through an open window, the person who opened the window might be considered responsible. 
However, when considering the technical system consisting of the robot, then there are multiple entities that might be responsible.

First, the owner of the robot might have used the robot against its specifications, \eg, if it might only be used in rooms with no fragile goods around. If, however, the robot was meant to be used in rooms with vases in them, then the manufacturer may be responsible for not adding a \enquote{vase detection mechanism}. In such a case it might be of further importance \emph{which} parts of the robot caused the destruction, \ie which (combination of) software or hardware module(s), eventually assigning responsibility to the provider of the corresponding module(s) or their integrator. Should the robot have been compromised by an attacker, then the  attacker is the root cause of the destruction. However, in many real world cases it is impossible to trace the attacker and assign blame to him. Thus, the question remains whether the manufacturer developed the robot in correspondence with state-of-the-art security practices. If this is not the case, then the manufacturer might be held responsible. 

In order to simplify the process of finding causes and assigning blame, an accountability system should be augmented with a causal reasoning system that helps an investigator parse the evidence created by logs. It should offer explanations and highlight possible causes and suspects.

\subsection{Liability}

These causal explanations can then be used to determine the legal liability for an event. Legal proceedings cannot be mechanized and will always require people to judge their peers.  Aiding the legal process is the ultimate goal of any  accountability system. Legal liability will not always be straightforward to determine. In the case of an malicious adversary hijacking the robot and breaking the vase, the owner of the robot might be to blame for not patching the robot; alternatively, the manufacturer might be to blame for not securing the robot better. In extremis, these questions will be answered in a legal trail. In such a trail the accountability mechanism should provide evidence that can be interpreted by the judges themselves or at least aid expert witnesses in their role as interpreters. 

\medskip
\noindent
Based on these considerations, \Cref{sec:model} presents our understanding of an accountability model.

\section{An Accountability Model}
\label{sec:model}


\begin{figure}[tb]
  \centering
   \includegraphics[width=0.5\textwidth]{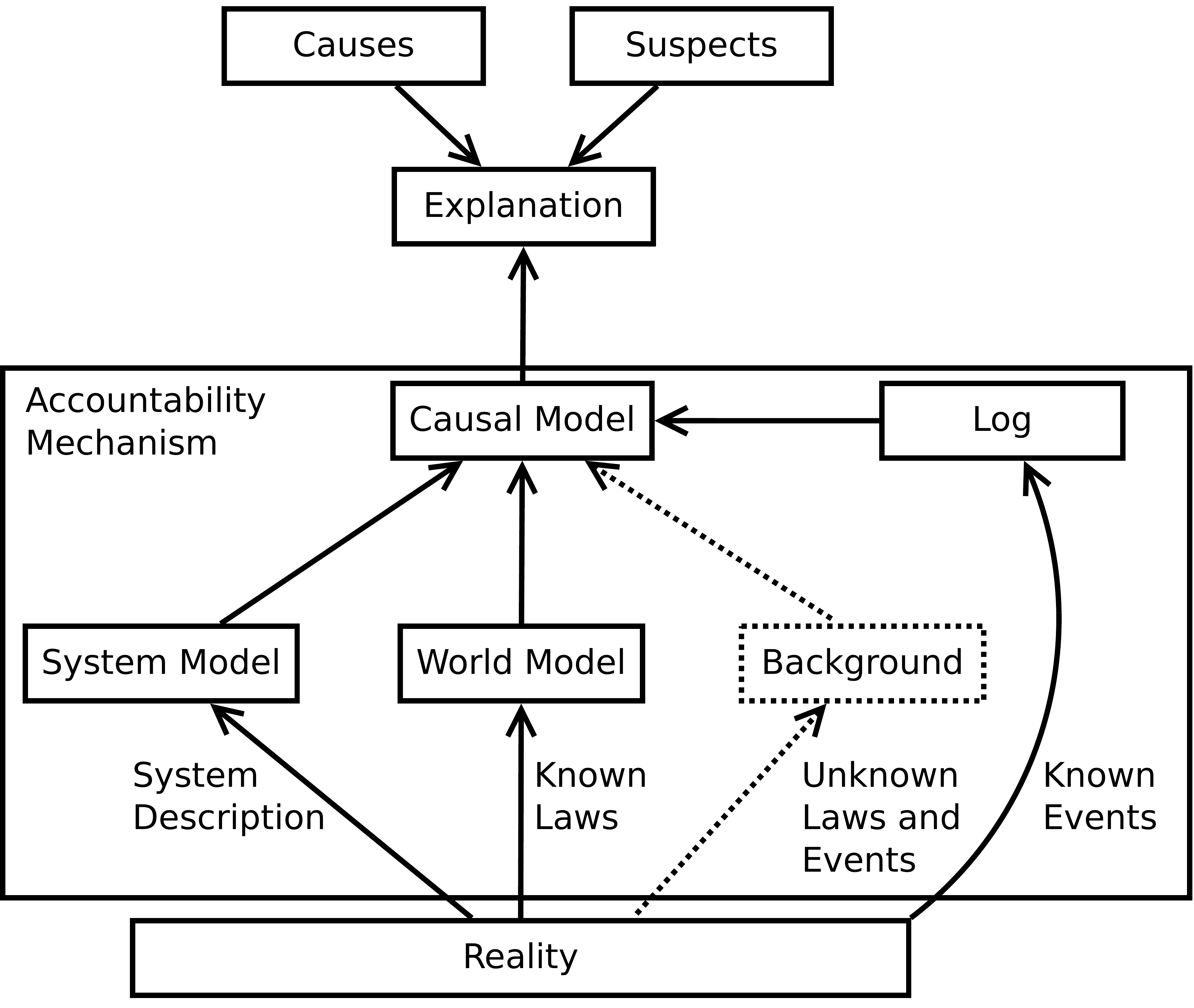}
  \caption{Our high level overview of accountability: Arrows read as \enquote{is represented in}. }
   \label{fig:high_level_model}
\end{figure}

As motivated in \Cref{sec:introduction,sec:preliminaries}, accountability mechanisms ought to explain why a certain behavior was (not) observed in the real world. Hence, our model aims to infer such explanations from real world observations. At the core of our model of accountability lies the \emph{causal model} in the tradition of Halpern and Pearl as described in \Cref{sec:relwork}. 
 \Cref{fig:high_level_model} depicts our high level understanding of such an accountability mechanism. The \emph{causal model}, which is detailed in \Cref{sec:causal-model},  will use knowledge about reality in the form of \emph{laws} and \emph{facts} to provide \emph{explanations}. The knowledge originates from four different sources: \begin{enumerate*}[(i)]
 \item the \emph{system model} represents (technical) knowledge of our system, 
 \item the \emph{world model} represents independent laws of the real world,
 \item the \emph{log} represents facts (or events) that have happened, and finally
 \item the \emph{background} represents everything we do not know
\end{enumerate*}.
While modeling the background might seem dubious at first, it makes explicit that there are events and causes that have not been thought of. Most importantly, this background will affect the \emph{causal model} by causing spurious relationships.

\subsection{World Model and System Model}
\label{sec:world-model}
\label{sec:system-model}

The world model captures \enquote{universal rules} that hold regardless of the system being modeled. Such rules include gravity, the knowledge that objects cannot move faster than the speed of light, or the idea that time moves forward. In contrast, a (technical) system model captures rules that (are expected to) hold within the observed (technical) system. Such rules include knowledge about the technical system altogether, its single hardware or software components, or the environment it acts in. In our example, such knowledge might include the robot's maximum speed, its battery's capacity, or the knowledge about the presence of a window in the room to be cleaned. Admittedly, distinguishing between the world model and the system model will not in all cases be trivial.
Ideally, both of these models should be \enquote{complete}, meaning that in no case any knowledge required for reasoning is missing.

Being aware that such completeness is likely to be unrealistic in practice, these models should be made as detailed as possible. The gap between reality and the corresponding models, \ie everything we do not know or have not considered upfront, is called \emph{background} (\cf \Cref{fig:high_level_model}) or \emph{exogenous variables} and may still affect the causal model, \eg, via confounding (\ie cause spurious relationships). Intuitively, the more knowledge is considered in the world model and the system model, the better will the quality of the reasoning provided by the causal model be. The fact that initial world models and system models are likely to miss relevant laws, requires them to be extendable/updatable.

\subsection{Causal Model and Reasoning}
\label{sec:causal-model}
The causal model is an \enquote{instance} of the world. It incorporates the (\enquote{always-true}) rules from the world and system model (\cf \Cref{sec:system-model}) and logs, \ie recordings of events of actual system runs (\cf \Cref{sec:prelim-logs}). As described, one  limitation of such a log-based approach is that only expected and observable (which is subject to the logging mechanism being used) events can be logged. Unexpected events will not be logged and enter the causal model as background variables. To improve the logs over time, we require that the logging mechanism can be updated during the lifetime of the system. 

We base the construction of causal diagrams on Pearl \cite{pearl2009causality} and use the following procedure to construct the causal diagrams for our instantiations in \Cref{sec:instantiations,sec:platform}:
\begin{enumerate}
	\item Draw a node for every fact in the log.
    \item Connect all nodes that have an observed causal relationship  with a solid directed edge. These are the causal relationships we have observed according to the logs.
    \item Connect all nodes whose causal relationship follows from the laws in the world and system model with a dashed directed edge. These are the causal relations that should have occurred. 
\end{enumerate}

Once the causal model has been constructed, we can use it to reason about what has happened in reality. Unsurprisingly, this \emph{reasoning} is then based on the world model, the system model, and the provided system logs. Eventually, such reasoning will generate \emph{explanations} consisting of \emph{causes} and possible \emph{suspects} (\cf \Cref{fig:high_level_model}). Causes are reasons for an event to happen, while suspects are concrete entities, \ie persons or organizations, that are to blame. To date, we are unsure whether, and, if so, in which cases, such reasoning and blaming can be fully automated; the possible degree of automation depends to a large degree on the available knowledge base. Hence, we also do not necessarily expect our system to find unique causes. Instead, our primary goal is to create a list of \enquote{candidate} causes and suspects, aiding humans in understanding the system's behavior. For example, we envision the user of this information to be a judge trying to settle a case or a developer tracking down a bug. We provide concrete examples for such reasoning in \Cref{sec:instantiation-security-causal,sec:instantiation-security-explanation}.



\section{Instantiations}
\label{sec:instantiations}

This and the following section show how our accountability model is able to capture security, privacy, and safety requirements. An instantiation for a security case focusing on an Unmanned Aerial Vehicle (UAV) is provided in \Cref{sec:instantiatio-security}, while \Cref{sec:instantiation-privacy} gives an intuition on how to initialize a privacy case based on the idea of \enquote{Information Accountability} \cite{weitzner2008information}. \Cref{sec:platform} shows how a safety case for our real-world research platform could look like. 


\subsection{Security}
\label{sec:instantiatio-security}
Our security use case features an Unmanned Aerial Vehicle (UAV) which enters restricted airspace, \eg, if it gets too close to an airport or military site. In such an event we want to know \emph{why} the UAV crossed into forbidden airspace. It could either have been steered there by the pilot, by a malicious third party or due to a technical failure, such as wrong location information. The accountability mechanism ought to help us discern the true cause. 

\subsubsection{System Model}
\label{sec:instantiation-security-system-model}

\begin{figure}[tb]
  \centering
   \includegraphics[width=0.6\textwidth]{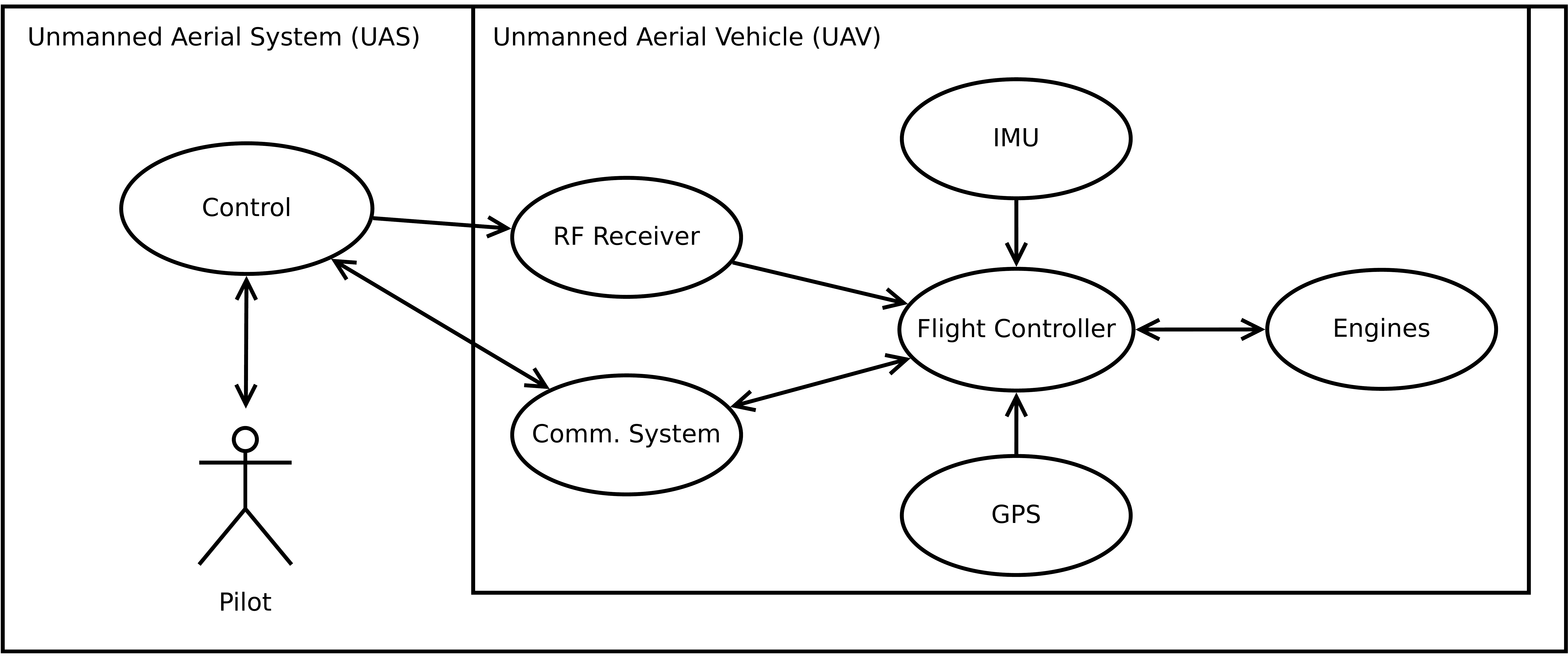}
  \caption{A high-level system model for an Unmanned Aerial System. Arrows represent information and command flows.}
   \label{fig:uav_system}
\end{figure}
\begin{figure}[tb]
  \centering
   \includegraphics[width=0.4\textwidth]{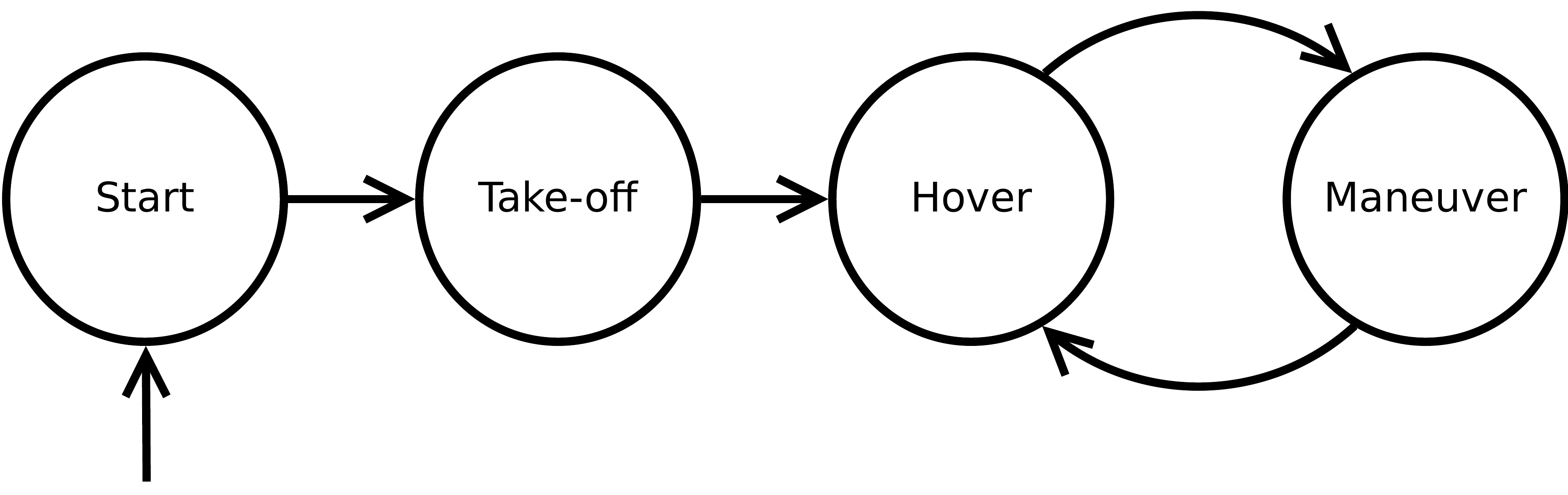}
  \caption{A state machine modeling the UAV's behavior.}
   \label{fig:uav_state}
\end{figure}

When modeling UAVs, usually two intertwined systems are considered: the Unmanned Aerial System (UAS), containing the pilot and any communication equipment, as well as the UAV (\Cref{fig:uav_system}). The UAV, as well as parts of the UAS, are technical systems and can be modeled accordingly. In our scenario we consider a human operator who interacts with a control system. We do not model the control system in detail, but imagine a remote control that sends commands to the UAV and receives telemetry data (\eg, height) from the UAV. The control system could also contain a receiver for a live video feed from the UAV to enable \enquote{first person flying}. A real-world example would be the Futaba controls used by AscTec\footnote{\url{http://www.asctec.de/en/uav-uas-drones-rpas-roav/asctec-falcon-8/mobile-ground-station/}} to control their UAVs. \Cref{fig:uav_state} depicts a simplified state machine that describes the UAV's behavior.

In the model of the UAV, the RF receiver will simply receive commands from the remote control and pass them on to the flight controller. The communication system can be a Wi-Fi connection to stream a live video feed, or a low bandwidth XBee link to transmit minimal telemetry data (\eg, height and GPS position). UAVs usually have at least two main sensors: a GPS receiver to determine their position in 3D space (often missing in cheap \enquote{toy UAVs}) and an Inertial Measurement Unit (IMU) that determines the aircraft's bearing and can be used to estimate a craft position and orientation (dead reckoning). Additionally, a UAV might be fitted with video cameras, barometers or laser scanners to independently measure their height above ground level, and a host of other sensors. All this data is usually processed at the central flight controller which then in turn controls the engines. In case of a quadrocopter, for example, the flight controller will keep the UAV hovering at a fixed position and height when it receives no commands from the remote control. 

From this model we can then infer basic causal connections within the system. We may, for example, want to figure out whether the pilot was responsible for steering the UAV into the forbidden airspace. If the pilot has not given any such steering command, our next \enquote{most likely} cause would be a malfunctioning remote control. If we can also rule out the remote control, we need to continue our investigation down the causality path, investigating whether any of the UAVs components malfunctioned. Lastly, also an attacker might have injected corresponding steering commands.

\subsubsection{World Model}
\label{sec:instantiation-security-world-model}
Among other things, the world model incorporates theories of gravity and motion, as well as a model for meteorological laws. Further, it might incorporate a database of GPS coordinates of restricted airspaces.

\begin{figure}[tb]
  \centering
\begin{lstlisting}
Timestamp   Component           Parent  Log Message
1           Pilot (P)           -       "I started the drone to take a video
                                        of the  nearby wood"
2           Control (C)         -       Control started
3           Flight Controller(FC)-      Startup and System self check
4           RF Receiver (RFR)   FC      online
5           IMU                 FC      online
6           Comm. System (CS)   FC      online
7           GPS                 FC      online
8           Engines (E)         FC      online
9           FC                  FC      all systems green       
10          P                   -       "starts UAV - full throttle"
11          C                   -       send control code START to UAV
12          RFR                 -       receive message START
13          FC                  RFR     START engines
14          E                   FC      Startup - Engines to 5V
15          C                   -       FULL THROTTLE (FT)
16          RFR                 -       received FT
17          FC                  RFR     FT to engines
18          E                   FC      Engines at 5V
19          GPS                 -       Current position is: 0/0
20          IMU                 -       Speed is: 5m/s Height:1m ...
...         ...                 ...     ...
600         P                   -       "I want to go left"
601         C                   -       send command GO LEFT to UAV
602         RFR                 -       received GO LEFT
603         FC                  RFR     received GO LEFT
604         FC                  FC      calculated new Engine speeds
605         E                   FC      set left engine 2.5V, others: 5V
606         GPS                 -       Current position is: 0/5
607         IMU                 -       Speed is:5m/s Height:25m...
\end{lstlisting}
\caption{Example log for the UAS. The parent field should be filled when messages can carry sender information and indicates which component issued the command.}
\label{fig:uav_log_structure}
\end{figure}

\subsubsection{Log}
\label{sec:instantiation-security-logs}
The recorded event log should offer as many details as possible, without being too costly in terms of runtime performance overhead and storage space. In this scenario, a comprehensive log could be structured similar to \Cref{fig:uav_log_structure}. This log shows in some detail how a UAV takes off and goes to an altitude of 25m. After reaching that altitude, the pilot steers the UAV to the left. Note that this log contains intentions of the pilot at timestamps 1, 10 and 600. These are non-technical events that are impossible to log using a technical accountability infrastructure. Hence, in this example a technical log was enriched with additional knowledge which was, \eg, acquired by asking the pilot (accepting that she might lie), or by inferring it from other witnesses. As even this simple example shows, creating \enquote{sufficiently complete} logs is a non-trivial task.

\begin{figure}[tb]
  \centering
\begin{lstlisting}
Timestamp   Component   Log Message
1           RC          Take-off command
2           UAV         All engines at full throttle
3           UAV         no RC commands; hover
4           RC          Go left command
5           UAV         Increase speed in left engine, keep others stable
6           RC          Go left command
7           UAV         already going left, keep going left
(repeat 6 and 7)
\end{lstlisting}
\caption{The pilot steering the UAV into the restricted airspace.}
\label{fig:uav_log1}
\end{figure}

To demonstrate the need to support different levels of abstraction, in addition we show two highly abstracted logs as they might have been created by a log parser after the UAV was found in restricted airspace. First, \Cref{fig:uav_log1} shows a log in which the pilot gives a clear command to go left, the direction of the supposed restricted airspace, at timestamps 4 to 7. The log depicted in  \Cref{fig:uav_log2}, in contrast, shows how the UAV goes left without any action from the pilot.

\begin{figure}[tb]
  \centering
\begin{lstlisting}
Timestamp   Component   Log Message
1           RC          Take-off command
2           UAV         All engines at full throttle
3           UAV         no RC commands; hover
4           RC          Go left command
5           UAV         Increase speed in left engine, keep others stable
6           RC          no command
7           UAV         already going left, keep going left
(repeat 6 and 7)
\end{lstlisting}
\caption{The pilot gives no command. The UAV should hover, yet still flies into the restricted airspace.}
\label{fig:uav_log2}
\end{figure}

\subsubsection{Causal Model}
\label{sec:instantiation-security-causal}

\begin{figure}[tb]
  \centering
   \includegraphics[width=0.7\textwidth]{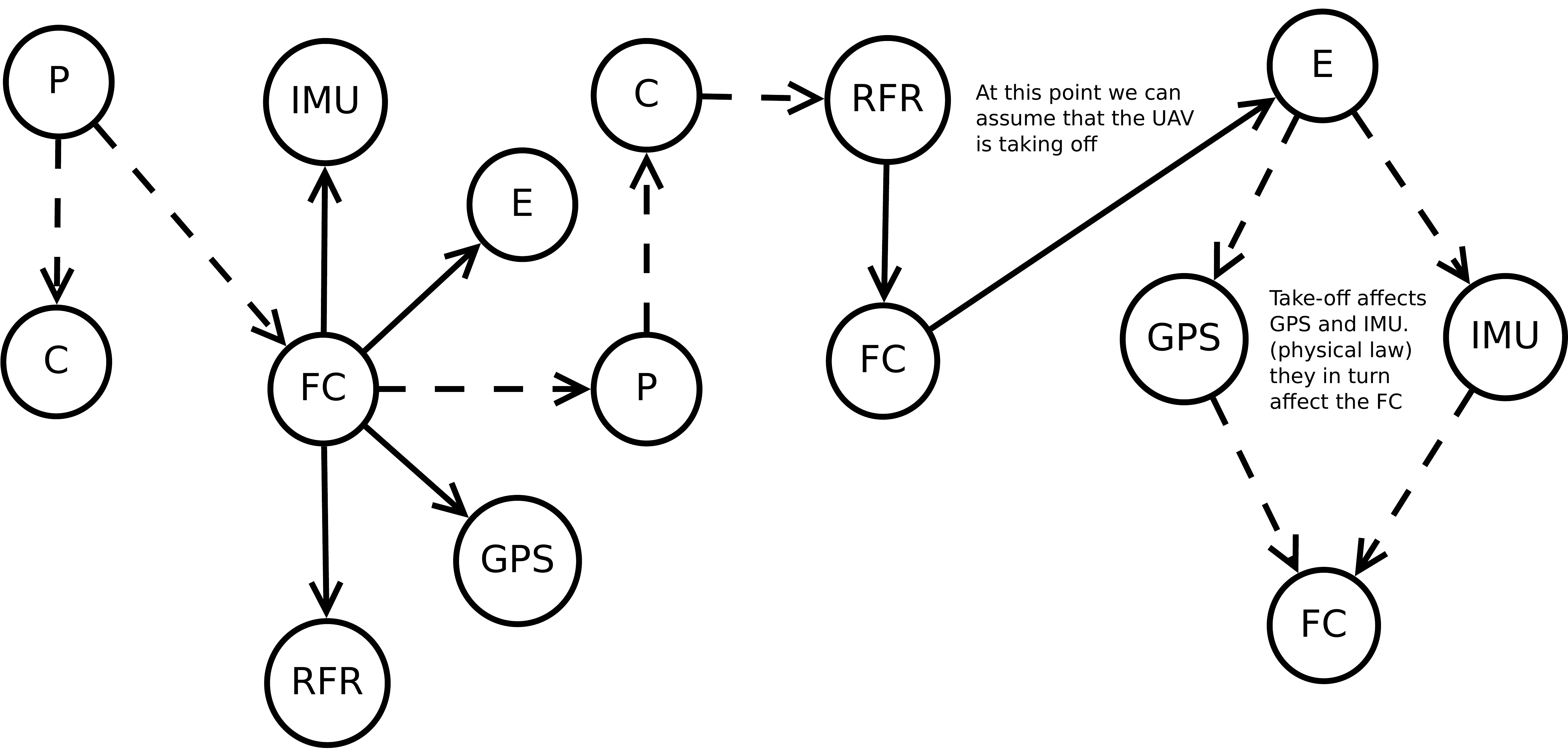}
  \caption{A causal model for the UAS example, encompassing the system model (\Cref{fig:uav_system}), the obtained logs (\Cref{fig:uav_log_structure}), and the world model (\Cref{sec:instantiation-security-world-model}).}
   \label{fig:uav_causal}
\end{figure}

The causal model of the UAV scenario, as depicted in \Cref{fig:uav_causal}, is obtained by combining the system model (\Cref{sec:instantiation-security-system-model}), the world model (\Cref{sec:instantiation-security-world-model}), and the obtained logs (\Cref{sec:instantiation-security-logs}). As described in \Cref{sec:causal-model}, log entries are represented as \emph{nodes}, while the system and world models' laws constitute the \emph{directed edges}. 

\begin{figure}[tb]
  \centering
   \includegraphics[width=0.7\textwidth]{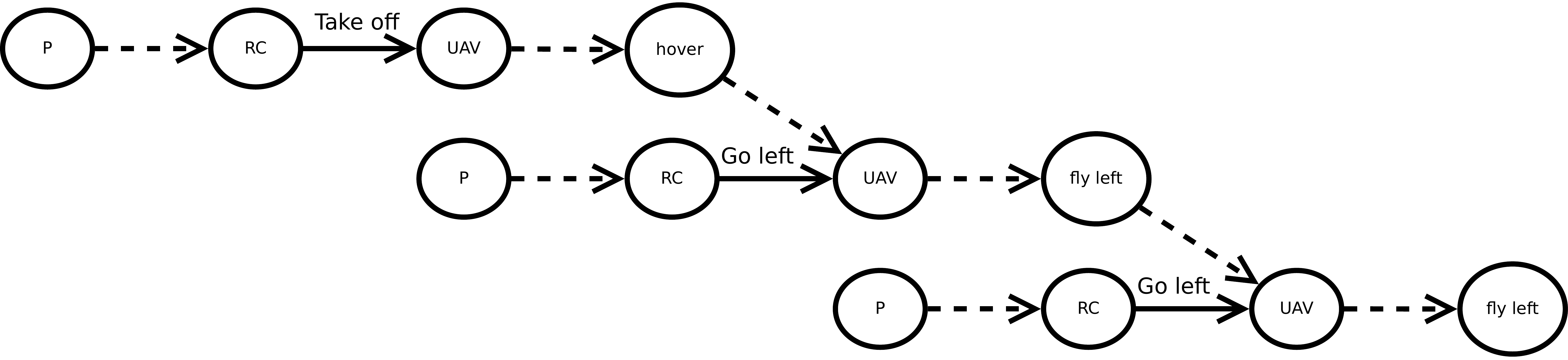}
   \caption{A causal model in which the pilot deliberately steers the UAV into an restricted airspace.}
   \label{fig:uav_causal1}
\end{figure}

\begin{figure}[tb]
  \centering
   \includegraphics[width=0.7\textwidth]{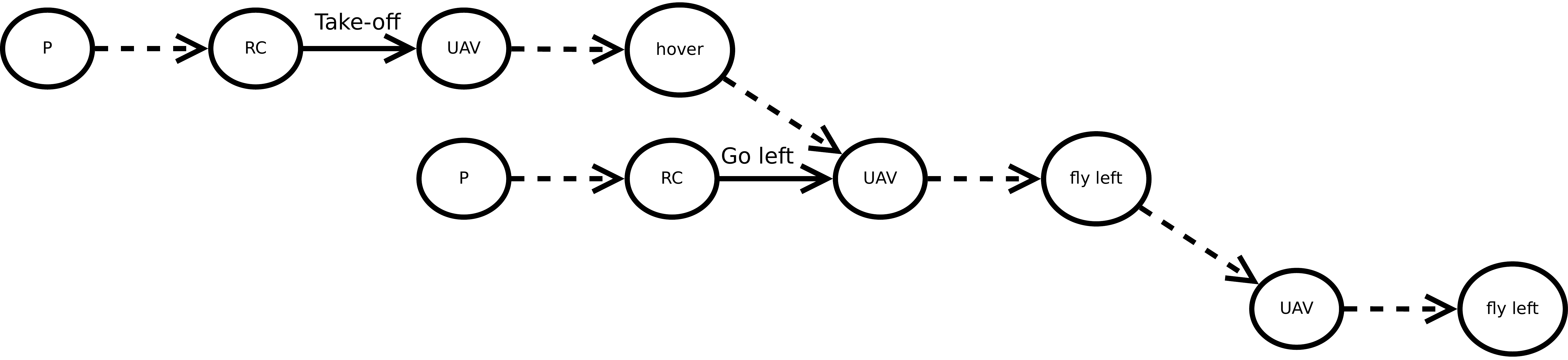}
  \caption{A causal model in which the pilot is not to blame; the fault lies with the UAV. }
   \label{fig:uav_causal2}
\end{figure}

Examining the model, we realize that it \enquote{ends} after the GPS and the IMU sent messages to the flight controller (Timestamp 20 in \Cref{fig:uav_log_structure}; right hand side of \Cref{fig:uav_causal}). If we correspond the nodes with the log (\Cref{fig:uav_log_structure}), we observe that the last message of the IMU indicated a height of 1m and a speed of 5m/s. This allows us to conclude that the UAV is indeed airborne. Following the arrows back to the origin, we can conclude that the pilot is the sole cause for take-off. Put as a counterfactual: had the pilot not started the UAV, it would never have taken off. If \Cref{fig:uav_causal} would lack an arrow from the pilot to either the control or the flight controller, the causal chain would end at the flight controller. We would then need to contemplate ways in which the flight controller could have started itself, \eg due to a faulty autopilot.

\Cref{fig:uav_causal1} depicts the (abstracted) causal model in which the pilot steers the UAV into the restricted airspace (the log is shown in \Cref{fig:uav_log1}). \Cref{fig:uav_causal2} is based upon the logs shown in \Cref{fig:uav_log2} and visualizes the fact that the UAV flew into the restricted airspace on its own accord and that the pilot did not give the respective commands. In the first case, \Cref{fig:uav_causal1}, we can see that the pilot repeatedly gives the \enquote{go left} command. If we consult our system model, we can see that without this command the UAV would just hover in the same spot. Thus we can assume that, had the pilot not given the command, the UAV would not have ended up in the restricted airspace. In the second example, depicted in  \Cref{fig:uav_causal2}, we can also see that the pilot started the UAV. So she is at least partly to blame. However, we can also see that she only gives a single \enquote{go left} command. Drawing on our understanding of the system, the UAV should have then hovered in place and not crossed into the restricted airspace. We would now need to investigate and find out why the UAV still kept on going left.

\subsubsection{Explanation}
\label{sec:instantiation-security-explanation}
These causal models can now be used to reason about causes for events and thus create \emph{explanations}. 
 We intend to create causal models based on the definitions by Halpern and Pearl \cite{halpern2005causes1,halpern2005causes2}, but have not yet formalized our approach on how to (semi-)automatically extract causes from such causal models.
Intuitively, we just need to follow the \enquote{arrows} back to their origin to find a set of root causes.  In the first example (\Cref{fig:uav_causal}) we can see that it was the pilot who started the UAV and thus she is on some level responsible for all events that follow. This can also be seen in  \Cref{fig:uav_causal1,fig:uav_causal2}: In both cases the pilot starts the UAV. The only difference is that in the latter example the pilot no longer gives commands to go left, the direction of the restricted airspace.

\subsection{Privacy}
\label{sec:instantiation-privacy}


In this section we provide the intuition of how to map the ideas for \enquote{Information Accountability} proposed by Weitzner \etal \cite{weitzner2008information} on our model. In their work they discuss ideas to move from the traditional \enquote{information hiding} approach for privacy to an \enquote{information accountability} approach. Instead of trying to keep information secret (which hardly ever works in the long term), they envision a world in which information is freely shared. However, an accountability system ensures that if information is accessed or used against the owner's intend, such abuse can be detected and traced back to the original perpetrator. To realize their idea, they propose an architecture that is comprised of three components: \begin{enumerate*}[(i)]
	\item \enquote{Policy Aware Transaction Logs},
    \item a \enquote{Policy Language Framework}, and
    \item \enquote{Policy Reasoning Tools}
\end{enumerate*}.

\emph{Policy Aware Transaction Logs} will \enquote{initially resemble traditional network and database transaction logs, but also include data provenance, annotations about how the information was used, and what rules are known to be associated with that information} \cite{weitzner2008information}. They can mostly be mapped to our notion of \emph{logs}. As they also include a notion of rules, however, some parts of these logs should naturally be mapped to either our system or world model. 

The \emph{Policy Language Framework} will be used to  \enquote{describ[e] policy rules and restrictions with respect to the information being used} \cite{weitzner2008information}. Just like we do for our world model, they require these policies to be shareable and extendable. They propose to use ontologies and semantic web technologies to merge these rules. This framework can easily be mapped to our notion of world and system model.

\emph{Policy Reasoning Tools} are then used to \enquote{assist users in seeking answers to questions [regarding their data]} \cite{weitzner2008information}. This concept can be mapped to our notion of a causal model.

\section{Research Platform}
\label{sec:platform}

To study how useful and capable the proposed accountability models and mechanism are in a real world setting, we plan to evaluate our results using a simple cyber-physical system. In line with our example, we chose a vacuum cleaning robot, thus being able to investigate the core problems of accountability for cyber-physical systems in a single use case. Additionally, we consider such a system a good proxy for more complex real-world cyber-physical systems since it
\begin{enumerate*}[(i)]
	\item moves in the physical world,
    \item interacts with people and objects,
    \item can operate autonomously, and
    \item can be remotely controlled
\end{enumerate*}.
In the future we plan to build an additional research platform based on a UAV.  

\begin{wrapfigure}{R}{8cm}
  \centering
   \includegraphics[width=8cm]{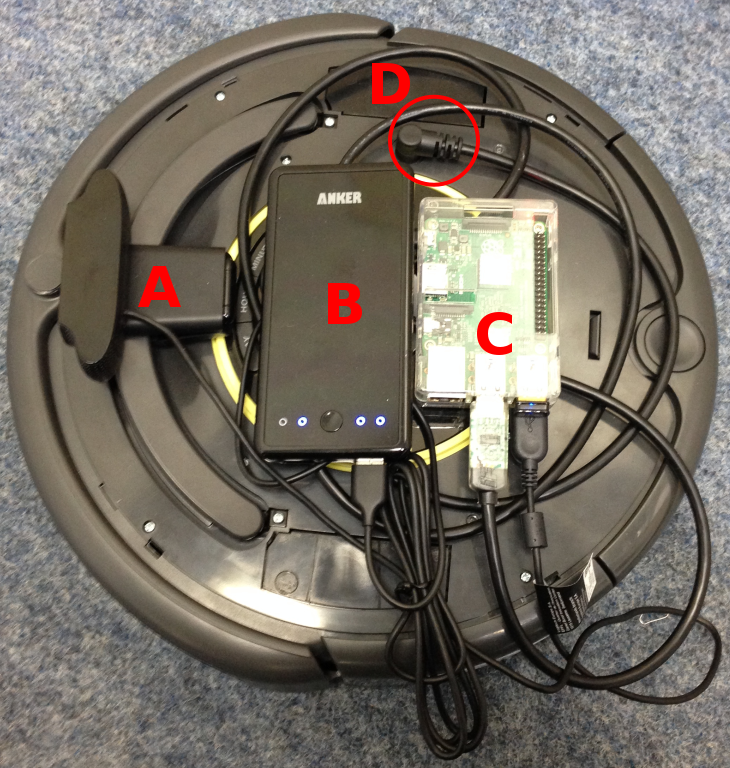}
  \caption{The Create robot: (A) Webcam, (B) Battery Pack, (C) RaspberryPi, (D) Serial-to-USB connector between the robot and the Pi.}
   \label{fig:roomba}
\end{wrapfigure}

\subsection{Technical Robot Description}

We base our platform on an iRobot Create (\Cref{fig:roomba}), the development version of the popular Roomba robotic vacuum cleaner\footnote{\url{http://wiki.ros.org/roomba_500_series}}. This robot features several bumper sensors that detect if the robot collides with an object, as well as cliff sensors that detect if the robot is about to fall off a cliff. Further, we  equipped the robot with a Full-HD webcam (including a H.264 video encoder) as well as a microphone. To be able to remote control the robot, we further added a battery-powered RaspberriPi and a USB WiFi adapter. Using these components, control commands as well as sensor data can be exchanged with any other computer. This setup allows us to simulate existing industrial service robots like the Aethon TUG\footnote{\url{http://www.aethon.com/tug/}}.

\begin{figure}[tb]
  \centering
   \includegraphics[width=0.9\textwidth]{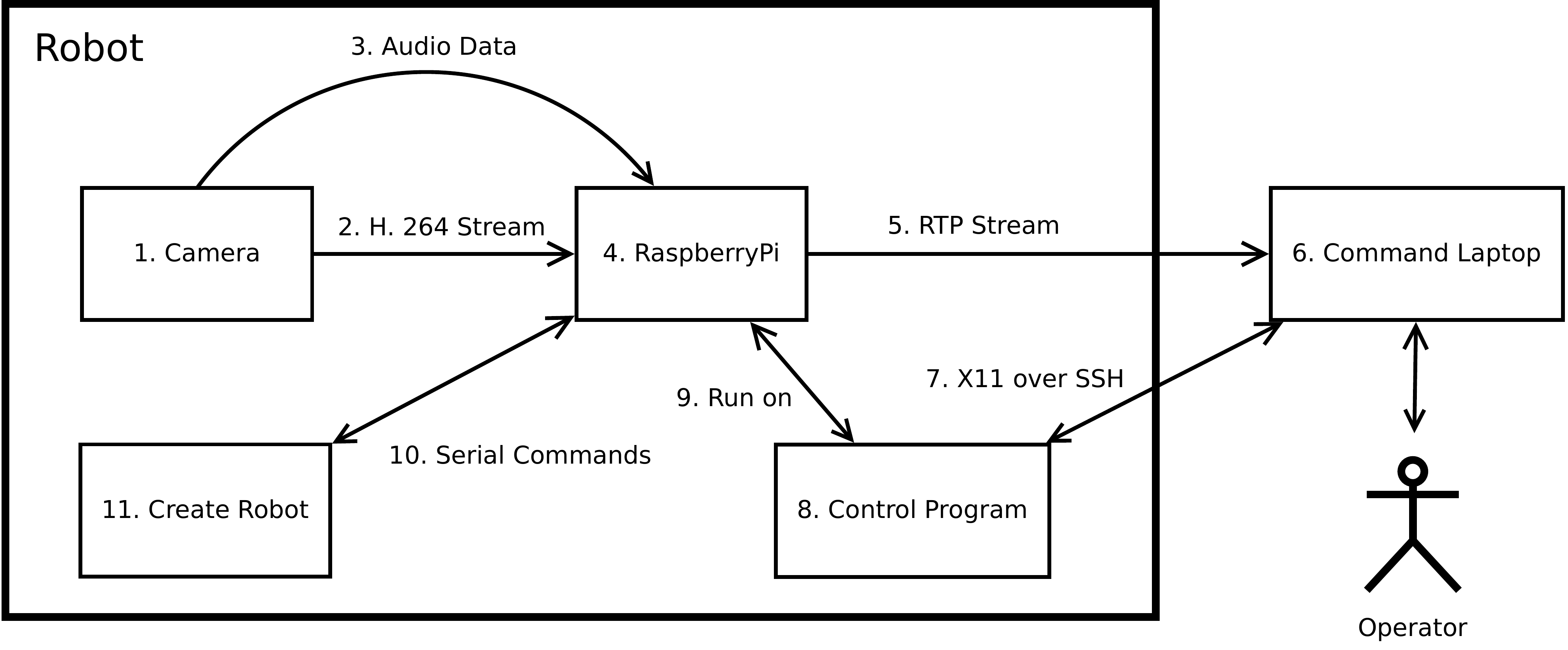}
  \caption{The high-level system model of our iRobot Create research platform.} 
   \label{fig:system_arch}
\end{figure}

\Cref{fig:system_arch} provides a high-level technical system diagram: The Logitech C920
camera (1) is used to record audio and video. The camera provides the video as
H.264 stream. The microphone is recognized as a normal Linux audio device. Video
and audio are transmitted to the RaspberryPi (4) via USB (2 and 3). To navigate
the robot (11), the video data is streamed to the command laptop (6) via an RTP
UDP Stream over WiFi (5). The robot (11) is controlled via serial commands sent
over an USB-to-serial cable (10) connected to the RaspberryPi (4). The serial
commands are sent by a Python script (8) that runs on the RaspberryPi (9) and is
displayed on the Command Laptop via X11-forwarding over SSH (7).

\subsection{Model Instantiation}
\label{sec:safety-model-init}
\begin{figure}
  \centering
  \begin{minipage}{0.35\textwidth}
    \centering
    \includegraphics[width=\textwidth]{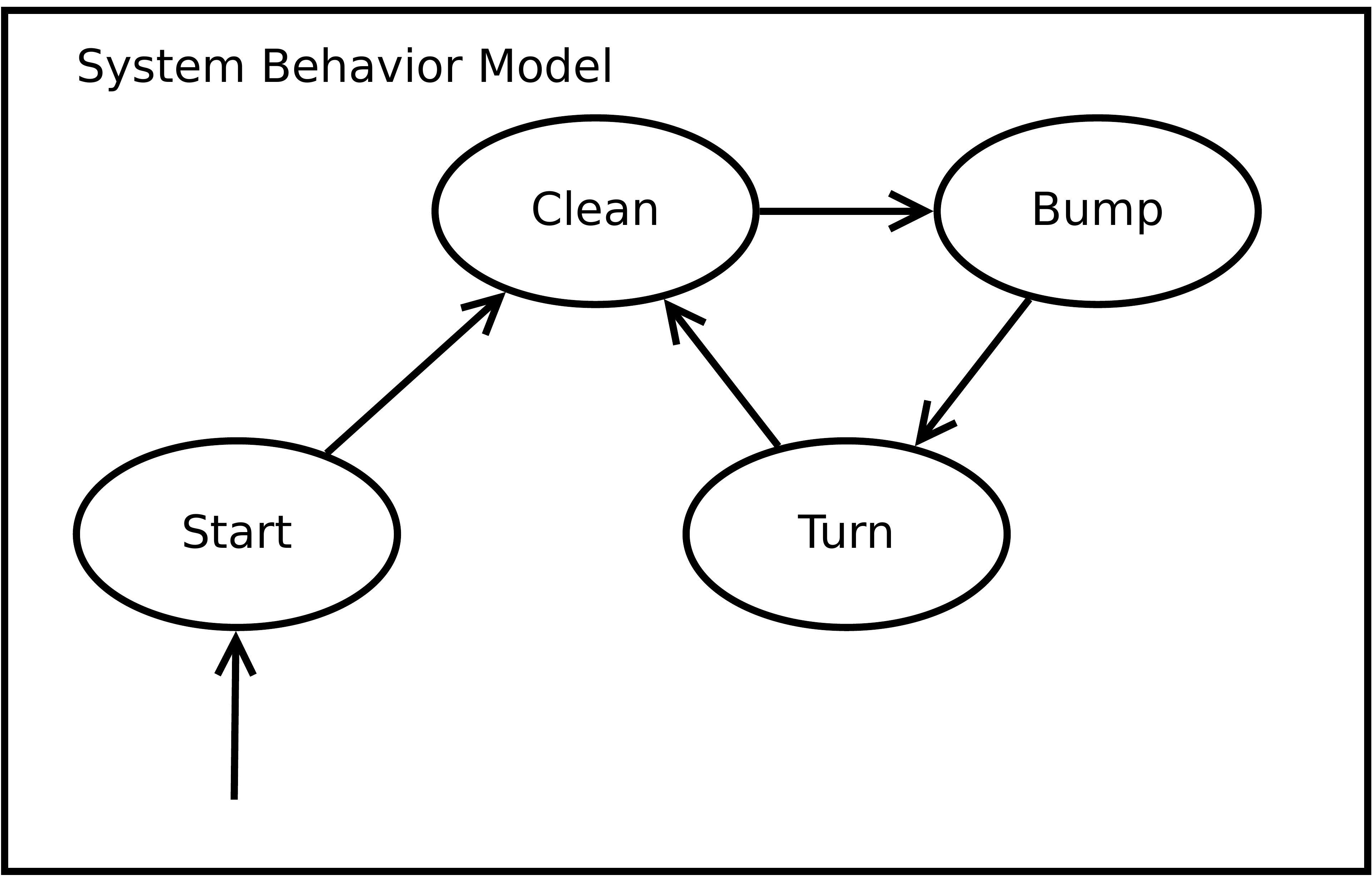}
    \caption{State machine that models the robot's behavior.}
    \label{fig:roomba_behavior}
  \end{minipage}\hfill
  \begin{minipage}{0.47\textwidth}
    \centering
\begin{lstlisting}
Time    Component   Message
0       Operator    "I started my Roomba"
3       Laptop      send start command to robot
5       Robot (R)   START
35      R           BUMP
70      R           BUMP
90      R           BUMP

\end{lstlisting}
    \caption{Example log for the cleaning robot.}
    \label{fig:roomba_log}
  \end{minipage}
\end{figure}

\begin{figure}[tb]
  \centering
   \includegraphics[width=0.5\textwidth]{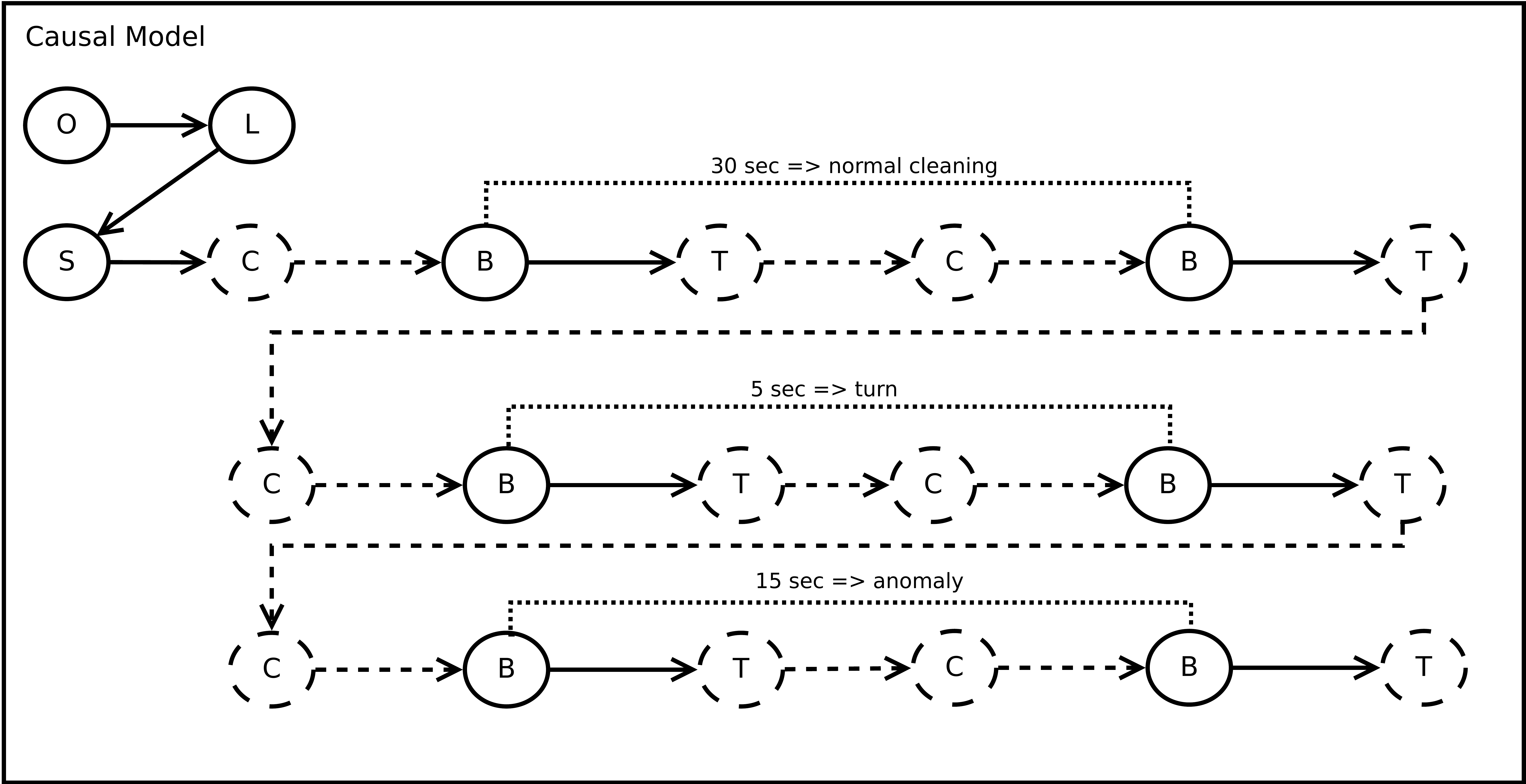}
  \caption{Causal model combining the system models (\Cref{fig:system_arch,fig:roomba_behavior}) with a log (\Cref{fig:roomba_log}).} 
   \label{fig:roomba_causal}
\end{figure}

In order to reason about the vase room (\Cref{sec:intro}), we extend our system model (\Cref{fig:system_arch}) with a simple system behavior model (\Cref{fig:roomba_behavior}). These models are then combined with the log (\Cref{fig:roomba_log}) to yield the causal model depicted in \Cref{fig:roomba_causal}. This causal model allows us to infer, that the operator (O) started the robot and that it could complete two \enquote{lanes} of cleaning without problems. When going down the third \enquote{lane}, it bumped against something after 15 seconds. By adding knowledge from the world model (each lane takes 30 seconds plus 5 seconds to turn), we can infer that this unexpected bump is an anomaly. If we now add the (supposed) fact that the windows where closed and no cat could enter the room, then we might conclude that it is highly likely that the robot did indeed break the vase. This conclusion can then be used to decide if the operator is to blame (because she did not remove the vase beforehand) or if the manufacturer is to blame (because they did not add a \enquote{vase detection mechanism}).

\FloatBarrier

\section{Conclusions and Future Work}
\label{sec:conclusions}

We believe that such accountability infrastructures will be useful in many fields and applications. They might even become a (legal) requirement in sensitive  or potentially dangerous domains (\eg, data processing or autonomous vehicles). Despite their obvious appeal, we have not seen a wide application of accountability mechanisms. In our opinion the reason for this is the complexity of the task: Causality has only recently been given a clear mathematical definition and reasoning about complex models is a hard problem (state explosion). 
We hope that designing and developing an accountability mechanism for a real world system will allow us to gain a deeper understanding of the problems and allow us to develop heuristics to keep the complexity of the model checking manageable. 

Our next steps will be to \enquote{fill the boxes} in our model of accountability with actual functionality, conduct the first real world experiments and gather real data. To this end we are currently in the process of building a system model that contains different layers of abstraction and join it with a basic model of our world (fortunately the world model for a cleaning robot can be kept simple). We are also evaluating different log styles and levels of granularity. As with the state explosion for models, too detailed logs supposedly cost too much performance. We are also not yet sure how to best incorporate external observations (\eg, the observations of a user) into our logs. 

We expect efficiency, both in terms of space and time, to be a core problem in any real-world accountability mechanism. If we save (storage) space by not logging some events, these events might be missing in the causal model. If, on the other hand, a logging mechanism slows down a system too much, its usability might be compromised.

Finally, we are also investigating non-technical aspects of accountability: \begin{enumerate*}[(i)]
\item How can accountability be incorporated into the juridical process, and
\item how can we ensure the accountability of an accountability system (second order accountability)
\end{enumerate*}? These questions are especially relevant in the context of autonomous vehicles, the Internet-of-Things, and the increasing autonomy of cyber-physical systems. We need to ensure that such systems confirm with current laws and regulations and, if that is not possible, suggest changes to these laws. To argue our case we need to ensure that accountability systems are safe and that, should something go wrong, we can explain why it went wrong and who should be held responsible. To solve these problems, we are confident that accountability infrastructures provide a promising approach.

\medskip\noindent
\textbf{Acknowledgments.}
We want to thank Eric Hilgendorf for first suggesting the vase example and our
colleagues Prachi Kumari and Kristian Beckers for valuable discussions on
accountability.
This research has been partially funded by the German Federal Ministry of 
Education and Research (BMBF) with grant number TUM: 01IS12057.

\bibliographystyle{eptcs}
\bibliography{bibliography}

\begin{thebibliography}{10}
\providecommand{\bibitemdeclare}[2]{}
\providecommand{\surnamestart}{}
\providecommand{\surnameend}{}
\providecommand{\urlprefix}{Available at }
\providecommand{\url}[1]{\texttt{#1}}
\providecommand{\href}[2]{\texttt{#2}}
\providecommand{\urlalt}[2]{\href{#1}{#2}}
\providecommand{\doi}[1]{doi:\urlalt{http://dx.doi.org/#1}{#1}}
\providecommand{\bibinfo}[2]{#2}

\bibitemdeclare{article}{avivzienis2004basic}
\bibitem{avivzienis2004basic}
\bibinfo{author}{A.~\surnamestart Avizienis\surnameend}, \bibinfo{author}{J.-C.
  \surnamestart Laprie\surnameend}, \bibinfo{author}{B.~\surnamestart
  Randell\surnameend} \& \bibinfo{author}{C.~\surnamestart Landwehr\surnameend}
  (\bibinfo{year}{2004}): \emph{\bibinfo{title}{{Basic Concepts and Taxonomy of
  Dependable and Secure Computing}}}.
\newblock {\sl \bibinfo{journal}{IEEE Transactions on Dependable and Secure
  Computing}} \bibinfo{volume}{1}(\bibinfo{number}{1}), pp.
  \bibinfo{pages}{11--33}, \doi{10.1109/TDSC.2004.2}.

\bibitemdeclare{}{ITIL}
\bibitem{ITIL}
\bibinfo{author}{\surnamestart {AXELOS}\surnameend} (\bibinfo{year}{2016}):
  \emph{\bibinfo{title}{{ITIL | AXELOS}}}.
\newblock \urlprefix\url{https://www.axelos.com/best-practice-solutions/itil}.

\bibitemdeclare{article}{dawid2000causality}
\bibitem{dawid2000causality}
\bibinfo{author}{Philip \surnamestart Dawid\surnameend} (\bibinfo{year}{2000}):
  \emph{\bibinfo{title}{{Causality Without Counterfactuals (With
  Discussion)}}}.
\newblock {\sl \bibinfo{journal}{J. Amer. Statist. Assoc}}
  \bibinfo{volume}{95}, \doi{10.1080/01621459.2000.10474210}.

\bibitemdeclare{inproceedings}{gossler2014blaming}
\bibitem{gossler2014blaming}
\bibinfo{author}{Gregor \surnamestart G\"{o}ssler\surnameend} \&
  \bibinfo{author}{L\u{a}cr\u{a}mioara \surnamestart
  A\c{s}tef\u{a}noaei\surnameend} (\bibinfo{year}{2014}):
  \emph{\bibinfo{title}{Blaming in Component-based Real-time Systems}}.
\newblock In: {\sl \bibinfo{booktitle}{Proceedings of the 14th International
  Conference on Embedded Software}}, \bibinfo{publisher}{ACM}, pp.
  \bibinfo{pages}{7:1--7:10}, \doi{10.1145/2656045.2656048}.

\bibitemdeclare{incollection}{gossler2014general}
\bibitem{gossler2014general}
\bibinfo{author}{Gregor \surnamestart G{\"o}ssler\surnameend} \&
  \bibinfo{author}{Daniel \surnamestart Le~M{\'e}tayer\surnameend}
  (\bibinfo{year}{2014}): \emph{\bibinfo{title}{A general trace-based framework
  of logical causality}}.
\newblock In: {\sl \bibinfo{booktitle}{Formal Aspects of Component Software}},
  \bibinfo{publisher}{Springer International Publishing}, pp.
  \bibinfo{pages}{157--173}, \doi{10.1007/978-3-319-07602-7\_11}.

\bibitemdeclare{book}{guagnin2012managing}
\bibitem{guagnin2012managing}
\bibinfo{author}{Daniel \surnamestart Guagnin\surnameend}
  (\bibinfo{year}{2012}): \emph{\bibinfo{title}{Managing privacy through
  accountability}}.
\newblock \bibinfo{publisher}{Palgrave Macmillan}, \doi{10.1057/9781137032225}.

\bibitemdeclare{article}{halpern2015}
\bibitem{halpern2015}
\bibinfo{author}{Joseph~Y. \surnamestart Halpern\surnameend}
  (\bibinfo{year}{2015}): \emph{\bibinfo{title}{{A Modification of the
  Halpern-Pearl Definition of Causality}}}.
\newblock {\sl \bibinfo{journal}{Proceedings of the 24th International Joint
  Conference on Artificial Intelligence}}, pp. \bibinfo{pages}{3022--3033}.
\newblock \urlprefix\url{http://arxiv.org/pdf/1505.00162}.

\bibitemdeclare{article}{halpern2005causes1}
\bibitem{halpern2005causes1}
\bibinfo{author}{Joseph~Y. \surnamestart Halpern\surnameend} \&
  \bibinfo{author}{Judea \surnamestart Pearl\surnameend}
  (\bibinfo{year}{2005}): \emph{\bibinfo{title}{Causes and Explanations: A
  Structural-Model Approach. Part I: Causes}}.
\newblock {\sl \bibinfo{journal}{The British Journal for the Philosophy of
  Science}} \bibinfo{volume}{56}(\bibinfo{number}{4}), pp.
  \bibinfo{pages}{843--887}, \doi{10.1093/bjps/axi147}.

\bibitemdeclare{article}{halpern2005causes2}
\bibitem{halpern2005causes2}
\bibinfo{author}{Joseph~Y. \surnamestart Halpern\surnameend} \&
  \bibinfo{author}{Judea \surnamestart Pearl\surnameend}
  (\bibinfo{year}{2005}): \emph{\bibinfo{title}{{Causes and Explanations: A
  Structural-Model Approach. Part II: Explanations}}}.
\newblock {\sl \bibinfo{journal}{The British Journal for the Philosophy of
  Science}} \bibinfo{volume}{56}(\bibinfo{number}{4}), pp.
  \bibinfo{pages}{889--911}, \doi{10.1093/bjps/axi148}.

\bibitemdeclare{}{COBIT}
\bibitem{COBIT}
\bibinfo{author}{\surnamestart {ISACA}\surnameend} (\bibinfo{year}{2016}):
  \emph{\bibinfo{title}{{COBIT 5: A Business Framework for the Governance and
  Management of Enterprise IT}}}.
\newblock \urlprefix\url{http://www.isaca.org/COBIT/Pages/default.aspx}.

\bibitemdeclare{inproceedings}{kusters2010accountability}
\bibitem{kusters2010accountability}
\bibinfo{author}{Ralf \surnamestart K\"{u}sters\surnameend},
  \bibinfo{author}{Tomasz \surnamestart Truderung\surnameend} \&
  \bibinfo{author}{Andreas \surnamestart Vogt\surnameend}
  (\bibinfo{year}{2010}): \emph{\bibinfo{title}{Accountability: Definition and
  Relationship to Verifiability}}.
\newblock In: {\sl \bibinfo{booktitle}{Proceedings of the 17th ACM Conference
  on Computer and Communications Security}}, \bibinfo{publisher}{ACM},
  \bibinfo{address}{New York, NY, USA}, pp. \bibinfo{pages}{526--535},
  \doi{10.1145/1866307.1866366}.

\bibitemdeclare{article}{papanikolaou2013cross}
\bibitem{papanikolaou2013cross}
\bibinfo{author}{Nick \surnamestart Papanikolaou\surnameend} \&
  \bibinfo{author}{Siani \surnamestart Pearson\surnameend}
  (\bibinfo{year}{2013}): \emph{\bibinfo{title}{A Cross-Disciplinary Review of
  the Concept of Accountability A Survey of the Literature}}.

\bibitemdeclare{book}{pearl2009causality}
\bibitem{pearl2009causality}
\bibinfo{author}{Judea \surnamestart Pearl\surnameend} (\bibinfo{year}{2009}):
  \emph{\bibinfo{title}{{Causality: Models, Reasoning and Inference}}},
  \bibinfo{edition}{2nd} edition.
\newblock \bibinfo{publisher}{Cambridge University Press},
  \doi{10.1017/CBO9780511803161}.

\bibitemdeclare{incollection}{pearson2009accountability}
\bibitem{pearson2009accountability}
\bibinfo{author}{Siani \surnamestart Pearson\surnameend} \&
  \bibinfo{author}{Andrew \surnamestart Charlesworth\surnameend}
  (\bibinfo{year}{2009}): \emph{\bibinfo{title}{Cloud Computing: First
  International Conference, CloudCom 2009, Beijing, China, December 1-4, 2009.
  Proceedings}}.
\newblock chapter \bibinfo{chapter}{Accountability as a Way Forward for Privacy
  Protection in the Cloud}, \bibinfo{publisher}{Springer Berlin Heidelberg},
  \bibinfo{address}{Berlin, Heidelberg}, pp. \bibinfo{pages}{131--144}.
\newblock \urlprefix\url{http://dx.doi.org/10.1007/978-3-642-10665-1_12}.

\bibitemdeclare{}{HIPAA}
\bibitem{HIPAA}
\bibinfo{author}{\surnamestart {U.S. Department of Health \& Human
  Services}\surnameend} (\bibinfo{year}{2016}): \emph{\bibinfo{title}{{Health
  Information Privacy | HHS.gov}}}.
\newblock \urlprefix\url{http://www.hhs.gov/hipaa/}.

\bibitemdeclare{article}{weitzner2008information}
\bibitem{weitzner2008information}
\bibinfo{author}{Daniel~J. \surnamestart Weitzner\surnameend},
  \bibinfo{author}{Harold \surnamestart Abelson\surnameend},
  \bibinfo{author}{Tim \surnamestart Berners-Lee\surnameend},
  \bibinfo{author}{Joan \surnamestart Feigenbaum\surnameend},
  \bibinfo{author}{James \surnamestart Hendler\surnameend} \&
  \bibinfo{author}{Gerald~Jay \surnamestart Sussman\surnameend}
  (\bibinfo{year}{2008}): \emph{\bibinfo{title}{Information Accountability}}.
\newblock {\sl \bibinfo{journal}{Commun. ACM}}
  \bibinfo{volume}{51}(\bibinfo{number}{6}), pp. \bibinfo{pages}{82--87},
  \doi{10.1145/1349026.1349043}.

\bibitemdeclare{article}{xiao2012survey}
\bibitem{xiao2012survey}
\bibinfo{author}{Zhifeng \surnamestart Xiao\surnameend},
  \bibinfo{author}{Nandhakumar \surnamestart Kathiresshan\surnameend} \&
  \bibinfo{author}{Yang \surnamestart Xiao\surnameend} (\bibinfo{year}{2012}):
  \emph{\bibinfo{title}{A survey of accountability in computer networks and
  distributed systems}}.
\newblock {\sl \bibinfo{journal}{Security and Communication Networks}},
  \doi{10.1002/sec.574}.

\end{thebibliography}

\end{document}